\documentclass[conference]{IEEEtran}
\IEEEoverridecommandlockouts

\usepackage{cite}
\usepackage{amsmath,amssymb,amsfonts}
\usepackage {algorithmic}
\usepackage{graphicx}
\usepackage[table]{xcolor}
\usepackage{textcomp}
\usepackage{xcolor}
\usepackage{booktabs} 
\usepackage{listings}
\usepackage{caption}
\usepackage{array}
\usepackage{tabularx}
\usepackage{multirow}
\usepackage{threeparttable}
\usepackage{subcaption}
\usepackage{placeins}
\usepackage{orcidlink}
\usepackage{pgfplots}
\pgfplotsset{compat=1.18}

\usepackage{tikz}
\usetikzlibrary{shapes, arrows.meta, positioning, fit, backgrounds, shadows, calc, decorations.pathmorphing}

\def\BibTeX{{\rm B\kern-.05em{\sc i\kern-.025em b}\kern-.08em
    T\kern-.1667em\lower.7ex\hbox{E}\kern-.125emX}}

\begin{document}

\title{From Design to Deorbit: A Solar-Electric Autonomous Module for Multi-Debris Remediation
}

\author{
  \IEEEauthorblockN{
    Om Mishra\IEEEauthorrefmark{1},
    Jayesh Patil\IEEEauthorrefmark{2},
    Sathwik Narkedimilli \orcidlink{0009-0001-9019-6283}\IEEEauthorrefmark{3},
    G Srikantha Sharma\IEEEauthorrefmark{4},\\
    Ananda S\IEEEauthorrefmark{5}, and
    Manjunath K Vanahalli\IEEEauthorrefmark{6}
  }
  \IEEEauthorblockA{
    \IEEEauthorrefmark{1}Department of Electronics \& Communication Engineering, Indian Institute of Information Technology Dharwad, India\\
  }
  \IEEEauthorblockA{
    \IEEEauthorrefmark{2}Department of Computer Science \& Engineering, Indian Institute of Information Technology Dharwad, India\\
  }
  \IEEEauthorblockA{
    \IEEEauthorrefmark{3}Department of Electrical \& Computer Engineering. National University of Singapore (NUS), Singapore\\
  }
  \IEEEauthorblockA{
    \IEEEauthorrefmark{4}Hindustan Aeronautics Limited (HAL), Bengaluru, India\\
  }
   \IEEEauthorblockA{
    \IEEEauthorrefmark{5}U R Rao Satellite Center, ISRO, Bengaluru, India\\ 
    }
  \IEEEauthorblockA{
    \IEEEauthorrefmark{6}Department of Data Science and Artificial Intelligence, Indian Institute of Information Technology Dharwad, India\\ 
  }
  
  \vspace{0.5em}
  \IEEEauthorblockA{
  Emails: 23bec035@iiitdwd.ac.in; 23bcs057@iiitdwd.ac.in; sathwik.narkedimilli@ieee.org; \\halseaking@gmail.com; ananda@ursc.gov.in;  manjunathkv@iiitdwd.ac.in
  }
}

\maketitle

\begin{abstract}
The escalating accumulation of orbital debris threatens the sustainability of space operations, necessitating active removal solutions that overcome the limitations of current fuel-dependent methods. To address this, this study introduces a novel remediation architecture that integrates a mechanical clamping system for secure capture with a high-efficiency, solar-powered NASA Evolutionary Xenon Thruster (NEXT) and autonomous navigation protocols. High-fidelity simulations validate the architecture's capabilities, demonstrating a successful retrograde deorbit from 800 km to 100 km, $<10$ m position Root Mean Square Errors (RMSE) via radar-basedExtended Kalman Filter (EKF) navigation, and a 93\% data delivery efficiency within 1 second using Delay/Disruption Tolerant Network (DTN) protocols. This approach significantly advances orbital management by establishing a benchmark for renewable solar propulsion that minimizes reliance on conventional fuels and extends mission longevity for multi-target removal.
\end{abstract}


\section{Introduction \& Related Works}

% current landscape
% problem statement
% how our research study fills the gap or the problem identified, and its relevance
% motivation
% contributions, implications, and application

\begin{figure}
    \centering
    \includegraphics[width=0.85\linewidth]{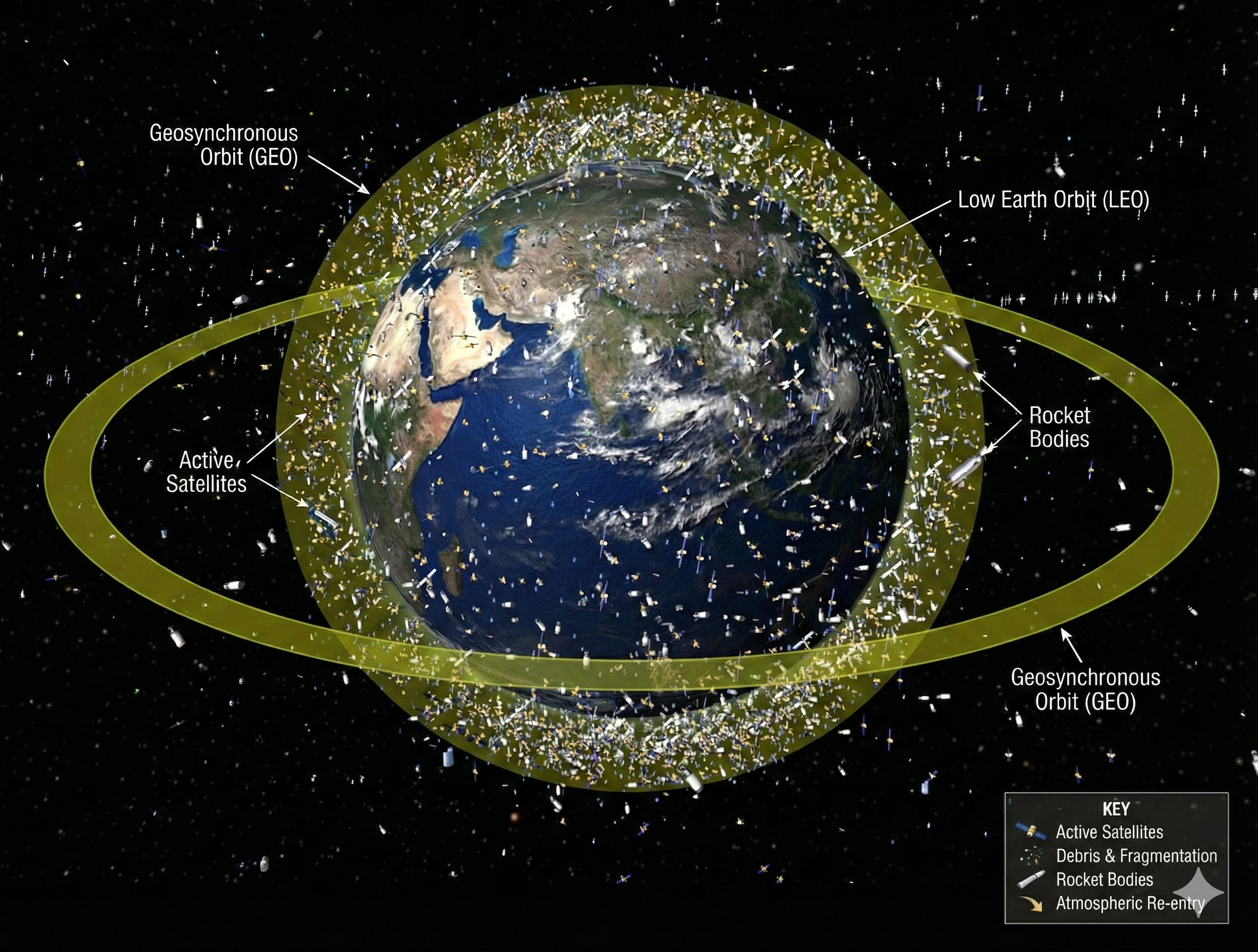}
    \caption{Schematic representation of Space debris distribution around Earth~\cite{esa2025kessler}.}
    \label{introimg}
\end{figure}

Since the dawn of space exploration, the rapid growth of commercial and scientific activities has led to a congested orbital environment, with an increasing population of operational spacecraft and growing space debris as schematically described in Fig.~\ref{introimg}. Over time, remnants from past missions and defunct spacecraft have accumulated in dense debris fields in key orbital regions, threatening the long-term safety and sustainability of space operations~\cite{smirnov2015space}. This buildup heightens the risk of collisions and cascading effects of Kessler Syndrome, endangering active satellites and critical infrastructure. Existing mitigation strategies, centered on prevention and disposal, cannot address the vast legacy debris, underscoring the urgent need for advanced active removal solutions~\cite{kessler1978collision}.

% Literature review

Orbital debris threatens spacecraft safety and mission continuity, motivating both contact and non-contact removal approaches, ranging from nets, robotic arms, and micropatterned dry adhesives (MDA) to pulsed lasers and ion-beam shepherding. A clear view of the environment is foundational. \textit{Liou et al.}~\cite{liou2008instability} surveyed debris populations, coupling statistical risk assessment with numerical simulations of size distributions and orbital decay to quantify collision probabilities and dense debris regions, while noting uncertainty for small fragments and evolving sources. These gaps motivate refined data-assimilative models and inform the focus of the present study on robust, scalable removal concepts tailored to uncertain, dynamic debris fields.

Contact-based methods continue to mature. \textit{Minghe Shan et al.}~\cite{shan2019contact} modeled net capture with penalty- and impulse-based contact dynamics using an Axis Based Bounding Box (AABB) detector and mass–spring nets. Both yielded comparable configurations (6\% difference), with penalty forces $< 6$~N and impulse peaks $< 5 \times 10^{-4}$~Ns, though stiffness and small time steps pose challenges. Gecko-inspired MDA offers strong, reversible bonding via van der Waals contact splitting and scalable fabrication~\cite{ben2022orbital}, yet its durability under radiation, thermal cycling, and varied surface conditions remains a concern. Harpoon systems demonstrate reliable anchoring~\cite{reed2013development} but face targeting, high-velocity impact, and damage risks. Robotic-arm Active Debris Capture and Removal (ADCR) advances zero-reaction maneuvers and robust control~\cite{zhang2022review}, but uncertainties, computational load, and integration challenges persist.

Non-contact strategies aim to reduce operational risk. \textit{Hang Xu et al.}~\cite{hang2025mission} optimize repeated removals with multi-spacecraft tasking via Greedy Randomized Adaptive Search Procedure with Large Neighborhood and Crossover Mechanisms (GRASP-LNCM) and a suboptimal search exploiting Lambert transfers, achieving high removal scores in Iridium-33 scenarios while contending with large search spaces, parameter sensitivity, and computational load. \textit{Yingwu Fang et al.}~\cite{fang2024low} model pulsed laser ablation of spherical Al alloy debris using Finite Element  Methods (FEM)  in COMSOL\textregistered~(laser power of 700–900~kW), reporting plume velocities up to 2.36~km/s (21~$\mu$s), surface temperatures near 9230~K (45~$\mu$s). After five passes (862 pulses), $\approx 627$~km =increases in semi-major axis, and the perigee drops from 256~km to 198.4~km is seen. Plasma–debris coupling fidelity and experimental validation remain open needs. Ion beam shepherding is a parallel non-contact avenue within this landscape.

Our study introduces a novel solar-powered ion thruster system integrated with an innovative mechanical clamping mechanism to enable the removal of multiple debris objects within a single mission. By combining ion propulsion efficiency with renewable solar energy, the system allows precise, controlled, and fuel-efficient deorbit maneuvers for debris of varying sizes and trajectories, reducing reliance on traditional propellants and supporting sustainable space operations~\cite{deluca2013active}~\cite{ebisuzaki2015demonstration}. In comparison, prior research spans statistical risk assessments (Liou \textit{et al.}), contact-based capture methods (Minghe Shan \textit{et al.}; Ben-Larbi \textit{et al.}), and non-contact approaches (Fang \textit{et al.}; Hang Xu \textit{et al.}), most of which target single-debris scenarios with propulsion and validation limitations (Zhang \textit{et al.}). Our scalable system addresses these gaps to enhance debris removal efficiency, sustainability, and orbital safety~\cite{bonnal2013active}. Key contributions include:

\begin{itemize}
   
\item A secure mechanical clamping mechanism for reliable debris capture and retention.
\item A high-efficiency ion propulsion system enabling controlled low-thrust deorbit maneuvers.
\item A solar-powered energy architecture supporting long-duration and fuel-efficient mission operations.
\item An integrated guidance, navigation, and control framework enabling autonomous sequential debris capture and deorbit missions.

\end{itemize}

The paper is structured as follows: Section-\ref{sec3} presents the system model, covering design, propulsion, and clamping mechanisms. Section-\ref{sec4} discusses simulation results and performance analysis, and Section-\ref{sec5} concludes with key findings, implications, and future research directions.

\section{System Design \& Architecture}\label{sec3}

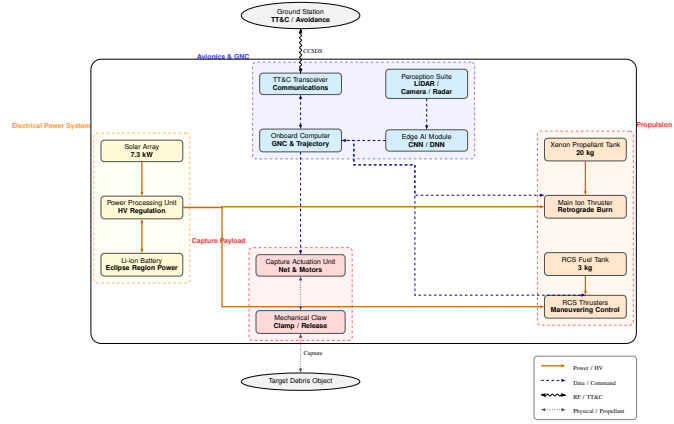
\begin{figure*}[htbp]
\centering
\resizebox{\columnwidth}{!}{%
\begin{tikzpicture}[
    node distance=1.4cm and 1.8cm,
    font=\sffamily\footnotesize,
    >={Latex[length=1.6mm]},
    % ---------- Base styles ----------
    base/.style={
        rectangle, rounded corners,
        draw=black!70, align=center,
        inner sep=5pt
    },
    avionics/.style={base, fill=cyan!15, text width=3.0cm},
    eps/.style={base, fill=yellow!20, text width=3.0cm},
    prop/.style={base, fill=orange!20, text width=3.0cm},
    payload/.style={base, fill=red!15, text width=3.3cm},
    ext/.style={
        ellipse, draw=black, thick,
        fill=gray!10, text width=3.2cm,
        align=center
    },
    % ---------- Link styles ----------
    power/.style={->, draw=orange!80!black, line width=1.5pt},
    data/.style={->, draw=blue!60!black, dashed, thick},
    phys/.style={<->, draw=black!70, thick, dotted},
    rf/.style={<->, draw=black, decorate,
        decoration={snake, amplitude=.4mm, segment length=2mm}}
]

% ================= EXTERNAL (TOP) =================
\node[ext] (ground) {Ground Station\\\textbf{TT\&C / Avoidance}};

% ================= AVIONICS =================
\node[avionics, below=1.8cm of ground] (ttc)
    {TT\&C Transceiver\\\textbf{Communications}};
\node[avionics, right=of ttc] (sensors)
    {Perception Suite\\\textbf{LiDAR / Camera / Radar}};
\node[avionics, below=of ttc] (obc)
    {Onboard Computer\\\textbf{GNC \& Trajectory}};
% AI Module included here
\node[avionics, right=of obc] (ai)
    {Edge AI Module\\\textbf{CNN / DNN}};

% ================= EPS =================
\node[eps, xshift=-6.5cm, below=1.8cm of obc] (ppu)
    {Power Processing Unit\\\textbf{HV Regulation}};
\node[eps, above=of ppu] (solar)
    {Solar Array\\\textbf{7.3 kW}};
\node[eps, below=of ppu] (batt)
    {Li-ion Battery\\\textbf{Eclipse Region Power}};

% ================= PROPULSION =================
\node[prop, xshift=6.5cm, below=1.8cm of ai] (ion)
    {Main Ion Thruster\\\textbf{Retrograde Burn}};
\node[prop, above=of ion] (tank)
    {Xenon Propellant Tank\\\textbf{20 kg}};
\node[prop, below=of ion] (rcs_tank)
    {RCS Fuel Tank\\\textbf{3 kg}};
\node[prop, below=0.8cm of rcs_tank] (rcs)
    {RCS Thrusters\\\textbf{Maneuvering Control}};

% ================= PAYLOAD =================
\node[payload, below=4.2cm of obc] (act)
    {Capture Actuation Unit\\\textbf{Net \& Motors}};
\node[payload, below=of act] (claw)
    {Mechanical Claw\\\textbf{Clamp / Release}};

% ================= EXTERNAL (BOTTOM) =================
\node[ext, below=1.6cm of claw] (debris)
    {Target Debris Object};

% ================= GROUP BOXES =================
\begin{scope}[on background layer]
    % Included 'ai' in the fit coordinates
    \node[
        fit=(ttc)(sensors)(obc)(ai),
        draw=blue!50, dashed, rounded corners,
        fill=blue!5, inner sep=8pt,
        label={[blue!80]north west:\textbf{Avionics \& GNC}}
    ] (grp_av) {};

    \node[
        fit=(solar)(ppu)(batt),
        draw=orange!60, dashed, rounded corners,
        fill=yellow!5, inner sep=8pt,
        label={[orange!80]north west:\textbf{Electrical Power System}}
    ] (grp_eps) {};

    \node[
        fit=(tank)(ion)(rcs_tank)(rcs),
        draw=red!60, dashed, rounded corners,
        fill=orange!5, inner sep=8pt,
        label={[red!80]north east:\textbf{Propulsion}}
    ] (grp_prop) {};

    \node[
        fit=(act)(claw),
        draw=red!60, dashed, rounded corners,
        fill=red!5, inner sep=8pt,
        label={[red!80]north west:\textbf{Capture Payload}}
    ] (grp_pay) {};

    \node[
        fit=(grp_av)(grp_eps)(grp_prop)(grp_pay),
        draw=black, thick, rounded corners=12pt
    ] (spacecraft) {};
\end{scope}

% ================= CONNECTIONS =================

% RF / TT&C
\draw[rf] (ground) -- node[right,font=\scriptsize]{CCSDS} (ttc);

% Data / Command (Updated logic)
\draw[data] (sensors) -- (ai);  % Sensors -> AI
\draw[data] (ai) -- (obc);      % AI -> OBC
\draw[data,<->] (obc) -- (ttc);
\draw[data] (obc.south) -- ++(0,-1.8) -- (act.north);

% Routing long data lines around components
\draw[data] (obc.east) -| ++(0.5,-1) -- ++(2.5,0) |- (ion.north west);
\draw[data] (obc.east) -| ++(0.5,-1) -- ++(2.5,0) |- (rcs.north);

% Power / HV
\draw[power] (solar) -- (ppu);
\draw[power,<->] (batt) -- (ppu);
\draw[power] (ppu.east) -- ++(1.6,0) |- (ion.west);
\draw[power] (ppu.east) -- ++(1.6,0) |- (rcs.west);

% Physical / Propellant
\draw[power] (tank) -- (ion);
\draw[power] (rcs_tank) -- (rcs);
\draw[phys] (act) -- (claw);
\draw[phys, thick] (claw) -- node[right,font=\scriptsize]{Capture} (debris);

% ================= LEGEND =================
% Position: Anchored to the South East of the Spacecraft box, shifted down
\node[
    draw=black!50,
    fill=white,
    rounded corners,
    anchor=north east,  % Anchor the top-right of the legend...
    minimum width=4.2cm, % Increased width
    minimum height=2.6cm % Increased height
] (legend) at ([yshift=-0.5cm]spacecraft.south east) {}; % ...to below the SE corner

% Legend Text
\node[
    anchor=north west,
    font=\scriptsize,
    align=left
] at ([xshift=1.5cm, yshift=-0.3cm]legend.north west)
{Power / HV\\[0.3cm] % Added vertical spacing
Data / Command\\[0.3cm]
RF / TT\&C\\[0.3cm]
Physical / Propellant};

% Legend Lines
\draw[power]
    ([xshift=0.3cm,yshift=-0.4cm]legend.north west) -- ++(1.0,0);
\draw[data]
    ([xshift=0.3cm,yshift=-1.0cm]legend.north west) -- ++(1.0,0);
\draw[rf]
    ([xshift=0.3cm,yshift=-1.6cm]legend.north west) -- ++(1.0,0);
\draw[phys]
    ([xshift=0.3cm,yshift=-2.2cm]legend.north west) -- ++(1.0,0);

\end{tikzpicture}}
\caption{Component-wise Architecture of the Space Debris Removal Module.}
\label{fig:architecture_rcs}
\end{figure*}

% Briefly describe the proposed system model
% Describe the tools used for simulation

The section presents the architecture of the proposed multi-debris remediation module, covering its propulsion, power, and guidance subsystems and their integration. It also outlines the configuration and key specifications that support the experimental simulations used to validate feasibility. The simulation parameters, based on established space-qualified hardware, are included to show the mission’s practical applicability.

\begin{figure*}[htbp]
    \centering
    \begin{subfigure}[t]{0.48\textwidth}
        \centering
    \includegraphics[width=0.85\textwidth]{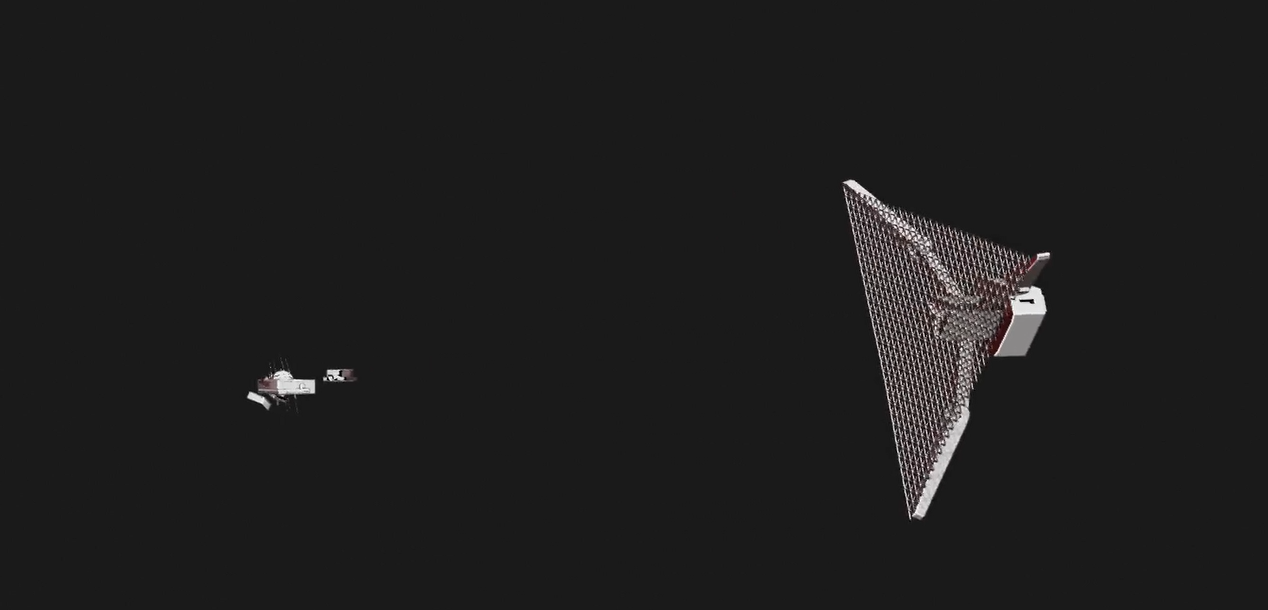}
        \caption{Device configuration for case~1}
        \label{fig:device1}
    \end{subfigure}
    \begin{subfigure}[t]{0.48\textwidth}
        \centering
    \includegraphics[width=0.75\textwidth]{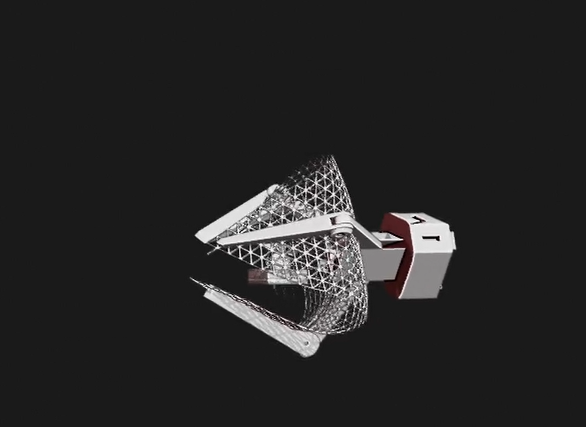}
        \caption{Device configuration for case~2}
        \label{fig:device2}
    \end{subfigure}
    \caption{Device layouts under different operating conditions.}
    \label{fig:device_results}
\end{figure*}

The architecture is designed to meet the key challenges of multi-debris removal: mission longevity, fuel efficiency, and autonomous operation. It combines a high-efficiency NEXT ion thruster with a solar power system to overcome propellant limits and support long-duration, sequential debris capture. A properly sized Li-ion battery maintains continuous thruster operation during LEO eclipse periods, addressing intermittent power. Advanced communication protocols and onboard processing enable autonomous navigation and reliable data handling. The module design is illustrated in Fig.~\ref{fig:device_results} under two operating configurations: (1) with the nets and claws in the open position for debris capture, and (2) with the claws in the closed position.

\begin{table*}[!t]
\centering
\caption{Summary of Proposed Solar-Powered Ion Propulsion Module Components.}
\label{tab:module_specs_integrated}
\scalebox{0.65}{
\begin{tabular}{l l p{2.8cm} l p{6cm}} 
\toprule
\textbf{Sub-system} & \textbf{Component} & \textbf{Key Specification} & \textbf{Value} & \textbf{Reference Value} \\
\midrule
\multirow{6}{*}{Propulsion} & Ion Thruster & Type & NEXT & Ref: NEXT data sheet \\
& & Thrust & 0.237 N & Ref: 0.237 N (NEXT at max power) \\
& & Specific Impulse (Isp) & 4100–4200 s & Ref: 4100–4200 s (NEXT) \\
& & Max Power Consumption & 6.9–7.3 kW & Ref: NEXT data sheet \\
\addlinespace
& Fuel (Propellant) & Type & Xenon & -- \\
& & Tank Capacity & 20 kg & Ref: 10–35 kg in recent missions \\
\midrule
\multirow{8}{*}{Power} & Primary Gen. & Type & Solar Array & -- \\
& (Solar Array) & Required Power (Max) & 7.3 kW & Sized for thruster + bus + recharge. Ref: 6–8 kW (Dawn, DART) \\
& & Specific Power (Basis) & 30 W/kg & Conservative, real-world value. Ref: 25–35 W/kg deployed (industry) \\
& & Calculated Mass & ~243 kg & (7,300 W / 30 W/kg). Ref: 230–250 kg for 7 kW in practice \\
\addlinespace
& Storage (EPS) & Type & Li-ion Battery & -- \\
& (Battery) & Required Energy & 4.1–5.7 kWh & Sized for LEO eclipse duration \\
& & Specific Energy (Basis) & ~170 Wh/kg & (with margins). Ref: 140–180 Wh/kg \\
& & Calculated Mass & ~33 kg & Ref: 30–40 kg, Li-ion \\
\midrule
\multirow{2}{*}{Spacecraft} & Structure / Bus & Dry Mass & 300 kg & Excludes propellant \& payload. Ref: 200–500 kg (deorbit probes) \\
& Payload (Target) & Example Mass & 100 kg & e.g., Debris object. Ref: Typically 50–150 kg class removed \\
\midrule
\multirow{2}{*}{Operational} & Mission Orbit & Profile & LEO (800-100 km) & Deorbit maneuver \\
& Context & Eclipse Duration & ~35 min & (per 95–100 min orbit) \\
\bottomrule
\end{tabular}
}
\end{table*}

The module's architecture, as shown in Fig.~\ref{fig:architecture_rcs}, is centered on a high-efficiency propulsion system designed for extended deorbit maneuvers. It uses a NEXT ion thruster, which provides a continuous thrust of 0.237 N with a high specific impulse (Isp) of 4100–4200 s. This system, which requires a maximum power of 6.9-7.3 kW, enables extended operation with a 20 kg Xenon propellant tank onboard. The spacecraft bus has a dry mass of 300 kg and is designed to capture and deorbit a 100 kg target payload. The mission profile involves a continuous deorbit maneuver from an 800-1000 km Low Earth Orbit (LEO), including an approximately 35-minute eclipse period per orbit during which the thruster must remain operational.

To meet the ion thruster's significant power requirements, the module integrates a large solar array capable of generating up to 7.3 kW. This array is designed assuming a conservative 30 W/kg specific power, yielding a calculated mass of approximately 243 kg. This primary generation system powers the thruster during sunlit phases and recharges the energy storage system. For continuous operation during the LEO eclipse, a 33 kg Li-ion battery pack at about 80\% depth of discharge (with reduced cycle life) is employed. This battery provides 4.1–5.7 kWh of energy at a specific energy of ~170 Wh/kg, ensuring the deorbit maneuver can proceed without interruption. Operating at this 80\% depth of discharge results in a projected lifespan of approximately 1000 sustainable cycles; consequently, the proposed mission's effective duration is estimated to be roughly three months. The sub-systems of the proposed module are summarized in the Table.~\ref{tab:module_specs_integrated}.

\begin{table*}[t]
\centering
\caption{Comprehensive Summary of Protocols Used in the Proposed Space Module Architecture.}
\label{tab:protocols_comprehensive}
\resizebox{0.65\textwidth}{!}{
  \begin{minipage}{\textwidth} 
    \begin{tabularx}{\textwidth}{ >{\raggedright}p{3.2cm} >{\raggedright}p{4.2cm} >{\raggedright\arraybackslash}X }
    \toprule
    \textbf{Protocol Category} & \textbf{Standard / Layer} & \textbf{Purpose in Proposed Module} \\
    \midrule
    
    \textbf{Long-Range TT\&C} &
    CCSDS TM/TC, Space Packets &
    Reliable uplink/downlink for commands, health data, and housekeeping telemetry during nominal cruise and debris-transfer phases. \\
    \addlinespace 
    
    \textbf{Physical Layers / RF Bands} &
    S-band, X-band, Ka-band, UHF, Optical &
    S-band for robust TT\&C; X/Ka for high-throughput payload downlink; UHF/Ka/Optical for short-range links (Proximity-1 or ISL). \\
    \addlinespace 
    
    \textbf{Bulk Data Transfer} &
    CCSDS CFDP (Class~1/2) &
    File-based transfer of compressed images, logs, and event recordings from onboard perception systems. Handles retransmission and link disruptions. \\
    \addlinespace 
    
    \textbf{Intermittent/Disrupted Links} &
    DTN Bundle Protocol (BP), BPSEC, HDTN &
    Store-and-forward delivery during ground passes or relay usage; ensures integrity and authentication of downlink bundles; supports high-rate payload dumps. \\
    \addlinespace 
    
    \textbf{Proximity Operations Link} &
    CCSDS Proximity-1 / UHF or Ka Short-Range Link &
    Low-latency communication during inspection, clamping, and debris capture. Supports deterministic control loops and rapid state exchange. \\
    \addlinespace 
    
    \textbf{Onboard Camera Data Handling} &
    CCSDS 122.0-B Image Compression (JPEG2000) &
    Lossless/lossy compression of frames, ROI prioritization, and efficient transfer of selected verification imagery. \\
    \addlinespace
    
    \textbf{Onboard Perception Pipeline} &
    CNN/DNN Edge Inference (TFLite/ONNX/PyTorch Mobile) &
    Local detection and tracking of debris; transmits only bounding boxes, confidence scores, and selected frames to reduce bandwidth load. \\
    \addlinespace
    
    \textbf{Radar Data Protocols} &
    ISO 23150 / AUTOSAR Radar Object List &
    Standardized object-list messages for range, velocity, and trajectory; time-aligned with camera data for fusion in the guidance loop. \\
    \addlinespace 
    
    \textbf{Internal Data Bus} &
    CAN / SpaceWire / UART (mission-dependent) &
    Transfers fused sensor states, actuator commands, and health reports between guidance, propulsion, and power subsystems. \\
    \addlinespace 
    
    \textbf{Safety \& Failsafe} &
    Heartbeat, Watchdog Timers, Fault Logs (CFDP/DTN) &
    Ensures safe retreat in the event of link loss, thruster faults, or failed capture; queues diagnostic bundles for later downlink. \\
    \addlinespace 
    
    \textbf{Service Prioritization (QoS)} &
    CCSDS Priorities, DTN Bundle Priorities &
    Manages bandwidth and link access by prioritizing safety-critical telemetry (e.g., attitude) $>$ detection metadata $>$ bulk payload dumps. \\
    \addlinespace 
    
    \textbf{Security} &
    BPSEC, TLS/DTLS (ground interfaces) &
    Ensures the integrity and authenticity of command traffic, sensor uploads, and bundle transfers. \\
    
    \bottomrule
    \end{tabularx}
  \end{minipage}
}
\end{table*}

A critical component of the module was the use of autonomous collision-avoidance logic. The mission profile operates on a dual-layer navigation strategy:
\begin{enumerate}
    \item \textbf{Global Avoidance:} The nominal trajectory is pre-planned based on ground-station data (e.g., TLE catalogs) to avoid known large satellites and tracked debris.
    \item \textbf{Local Avoidance:} Onboard sensors are tasked with detecting uncataloged, sub-centimeter debris that ground stations cannot track.
\end{enumerate}

The simulation tested the maneuvering model's ability to fuse these data sources. If the onboard perception system detects a previously unknown object on a collision course, the guidance algorithm is designed to trigger an immediate \textit{trajectory interrupt}. The system computes an optimal evasion maneuver to clear the threat, then autonomously re-optimizes the path to resume the deorbit mission.

The module employs a multi-layered communication architecture for robust control and data handling. Long-range Telemetry, Tracking, and Command (TT\&C) uses CCSDS (Consultative Committee for Space Data Systems) standards over S-band for reliability, with X/ka Bands reserved for high-throughput data downlink. To manage intermittent ground passes, the Delay/Disruption-Tolerant Networking (DTN) Bundle Protocol is used for store-and-forward delivery of bulk data, with the CCSDS File Delivery Protocol (CCSDS CFDP) handling the data. During critical proximity operations for inspection and capture, the system switches to a low-latency CCSDS Proximity-1 link. Internally, a high-speed bus like SpaceWire or CAN manages the transfer of fused sensor states and commands between subsystems.

To optimize bandwidth, the architecture relies heavily on onboard edge processing. CNN/DNN (Convolutional Neural Network/Deep Neural Network) inference engines locally detect and track debris, sending only metadata rather than raw sensor feeds. Radar data are standardized using ISO 23150 object lists, whereas camera data are compressed using CCSDS 122.0-B (JPEG2000). CCSDS priorities govern this data flow to ensure that safety-critical telemetry is prioritized over bulk file dumps. Fail-safe mechanisms, including heartbeat and watchdog timers, provide a safe fallback in the event of link loss or component faults. Security is maintained using BPSEC (Bundle Protocol Security) and TLS/DTLS (Transport Layer Security (TLS) and Datagram Transport Layer Security) to ensure the integrity and authenticity of all command traffic and data transfers. The protocols used in the proposed module are summarized in the Table.~\ref{tab:protocols_comprehensive}.

A comprehensive simulation campaign was conducted in MATLAB\textregistered\ and GMAT (General Mission Analysis Tool) to validate the proposed architecture, whose results and inferences are discussed in the next section. The analysis covered: (1) a high-fidelity GMAT simulation of the full low-thrust retrograde deorbit from 800 km to 100 km; (2) end-to-end EKF validation for the truth-sensor-estimator navigation chain with thruster perturbation correction; (3) verification of the radar-only relative navigation system for close-proximity operations; (4) evaluation of multi-hop DTN protocol performance and link reliability; and (5) numerical simulation of the low-thrust spiral trajectory.

\section{Results and Discussion}\label{sec4}

\begin{figure}[h]
    \centering
    \resizebox{0.8\columnwidth}{!}{%
        \includegraphics{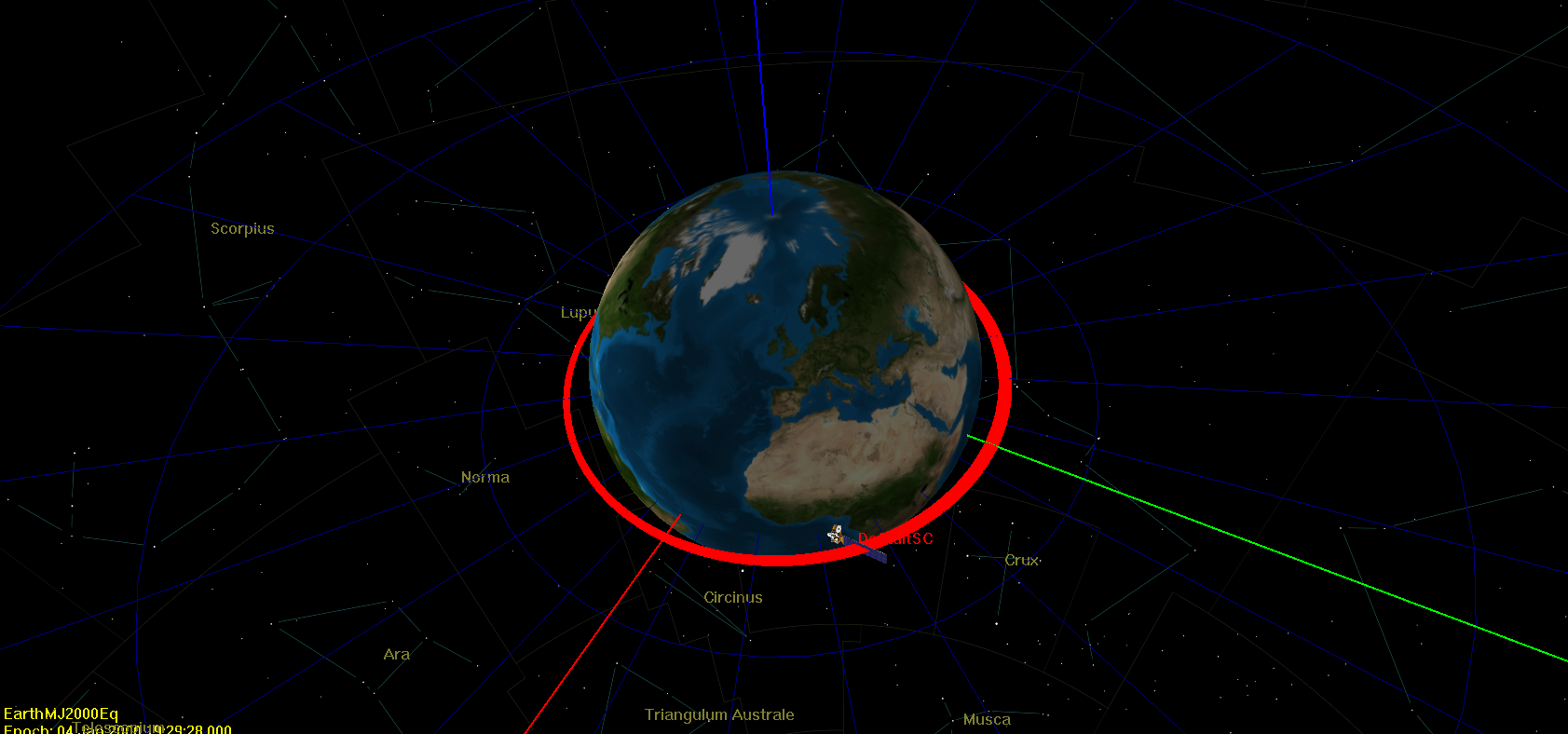}
    }
    \caption{GMAT simulation results under different operating conditions.}
    \label{fig:gmat_results}
\end{figure}

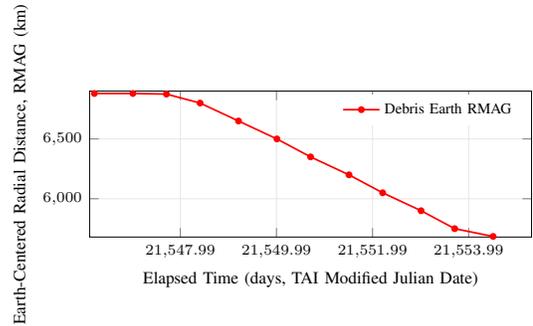
\begin{figure}[htbp]
    \centering
    \resizebox{0.8\columnwidth}{!}{%
    \begin{tikzpicture}
        \begin{axis}[
            width=\columnwidth,
            height=0.45\columnwidth,
            grid=both,
            grid style={gray!20},
            xlabel={Elapsed Time (days, TAI Modified Julian Date)},
            ylabel={Earth-Centered Radial Distance, RMAG (km)},
            xmin=21546.1, xmax=21555.3,
            ymin=5680, ymax=6900,
            ticklabel style={font=\scriptsize},
            label style={font=\footnotesize},
            scaled x ticks=false,
            xticklabel style={/pgf/number format/fixed},
            legend style={
                font=\scriptsize,
                at={(0.98,0.98)},
                anchor=north east,
                draw=none
            }
        ]
            \addplot[
                red,
                thick,
                mark=*,
                mark size=1.2
            ] coordinates {
                (21546.2,6880)
                (21547.0,6880)
                (21547.7,6875)
                (21548.4,6800)
                (21549.2,6650)
                (21550.0,6500)
                (21550.7,6350)
                (21551.5,6200)
                (21552.2,6050)
                (21553.0,5900)
                (21553.7,5750)
                (21554.5,5685)
            };
            \addlegendentry{Debris Earth RMAG}
        \end{axis}
    \end{tikzpicture}
    }
    \caption{Radial distance of debris relative to Earth over time.}
    \label{fig:debris_rmag}
\end{figure}

The validation of the proposed multi-space debris remediation architecture employed a dual-simulation framework, combining GMAT for high-fidelity orbital dynamics modeling and MATLAB for subsystem-level verification. As illustrated in Fig.~\ref{fig:gmat_results}, the GMAT analysis utilizes a physics-based truth model to capture the spacecraft’s gradual inward orbital decay from approximately 800~km to 100~km under sustained retrograde low-thrust acceleration. Fig.~\ref{fig:debris_rmag} presents the evolution of the Earth-centered radial distance (EarthMJ2000Eq) as a function of elapsed time expressed in International Atomic Time (TAI) Modified Julian Days. The altitude profile exhibits an initially near-flat phase, followed by a smooth, monotonic, and approximately linear decay, reflecting the cumulative reduction of specific orbital energy characteristic of continuous low-thrust maneuvers and confirming stable, controlled deorbiting to the termination altitude.

Unlike impulsive deorbit strategies, the continuous-thrust profile produces a smooth and sustained decay in orbital altitude, consistent with the long-duration dynamics of electric propulsion systems. By accounting for Earth-centered inertial propagation and thrust-induced orbital energy dissipation, the simulation captures the gradual nature of the maneuver over multiple orbital revolutions. Complementary to this mission-level truth reference, MATLAB-based simulations were used to evaluate the onboard estimation and communication subsystems, demonstrating the robustness of the overall mission architecture under varying operating conditions.

\begin{figure}
        \centering
    \includegraphics[width=0.45\textwidth]{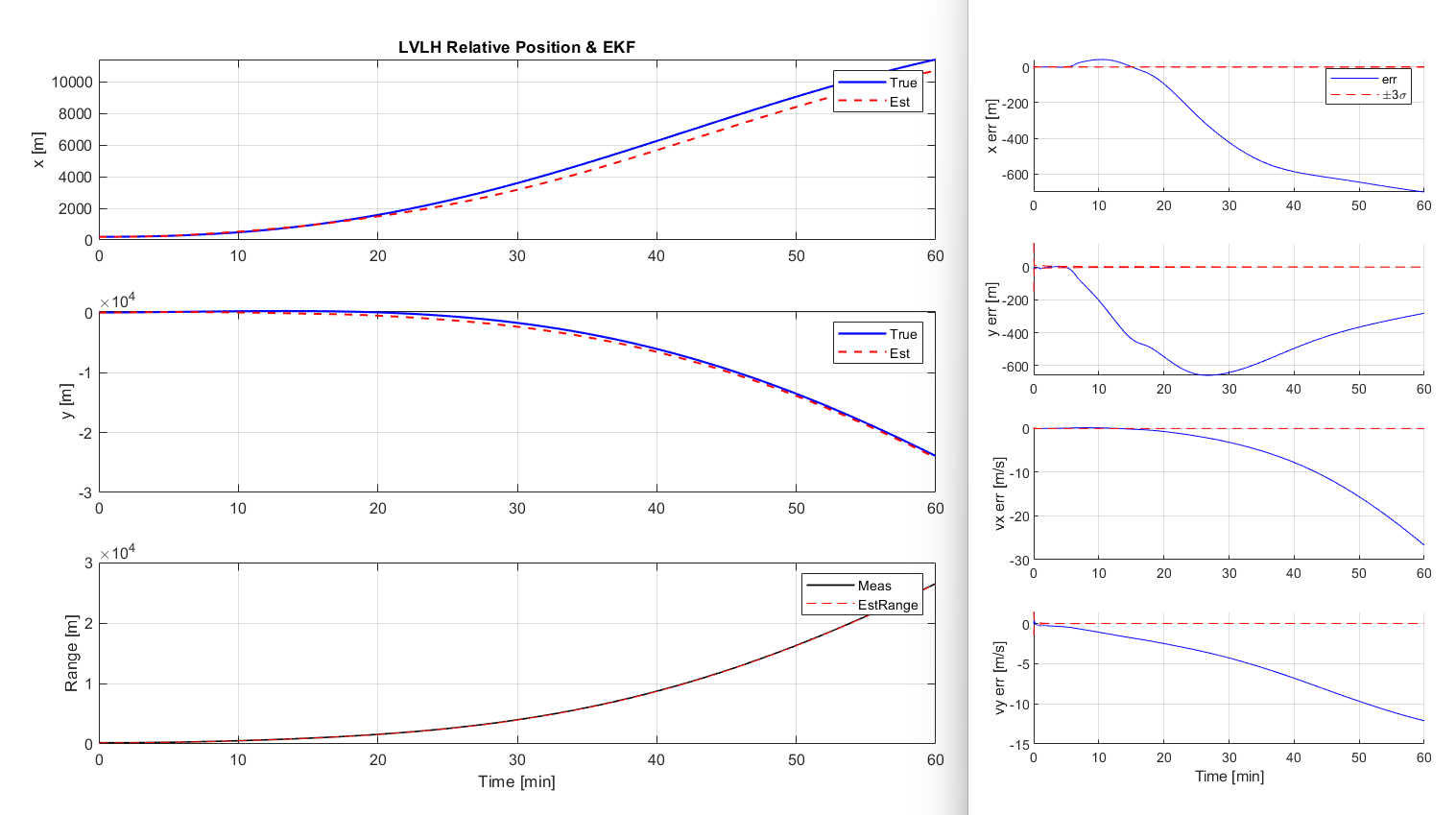}
        \caption{Buffer Backlog Evolution in the DTN Relay Architecture.}
        \label{lvlh-fig5}
\end{figure}

The performance of the onboard navigation system was evaluated by comparing the Extended Kalman Filter (EKF) estimates against the truth trajectory. Fig.~\ref{lvlh-fig5} illustrates the specific Local Vertical Local Horizontal (LVLH) relative position evolution. Over a 60-minute horizon, the radial separation grew from 200~m to over 10~km due to continuous 0.3~N tangential thrust. While the EKF successfully tracked this motion with an RMS position error of $10^2$~m, a gradual divergence was observed, attributed to the mismatch between the nonlinear thrust-perturbed truth and the linearized Clohessy-Wiltshire (CW) process model. However, in closer-proximity scenarios where relative motion is constrained (200~m to 1.4~km), the system achieved significantly higher precision. Here, the EKF maintained position errors within $\pm8$~m and velocity errors below 0.03~m/s, as confirmed by the tight covariance bounds. These results demonstrate that while radar-only navigation ($\sigma_r = 1$~m) offers strong observability and meter-level accuracy for inspection, long-duration spiraling requires thrust-aware process modeling to mitigate linearization biases.

\begin{figure}
        \centering
        \includegraphics[width=0.4\linewidth]{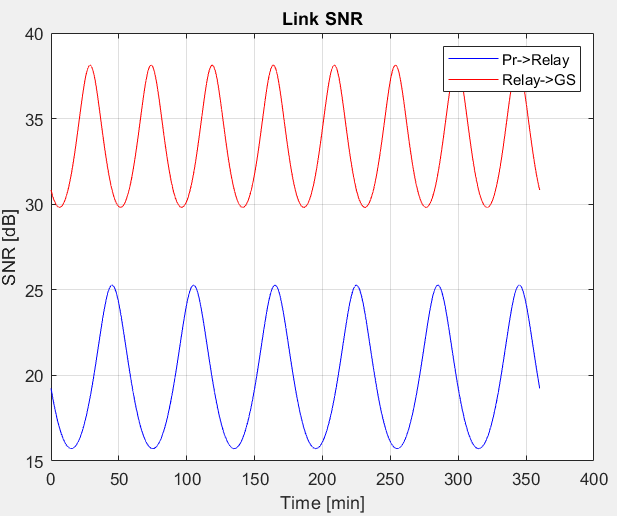}
        \caption{Link SNR vs Time (Primary→Relay and Relay→GS).}
        \label{linksnr-fig10}
\end{figure}

\begin{figure}
        \centering
     \includegraphics[width=0.45\textwidth]{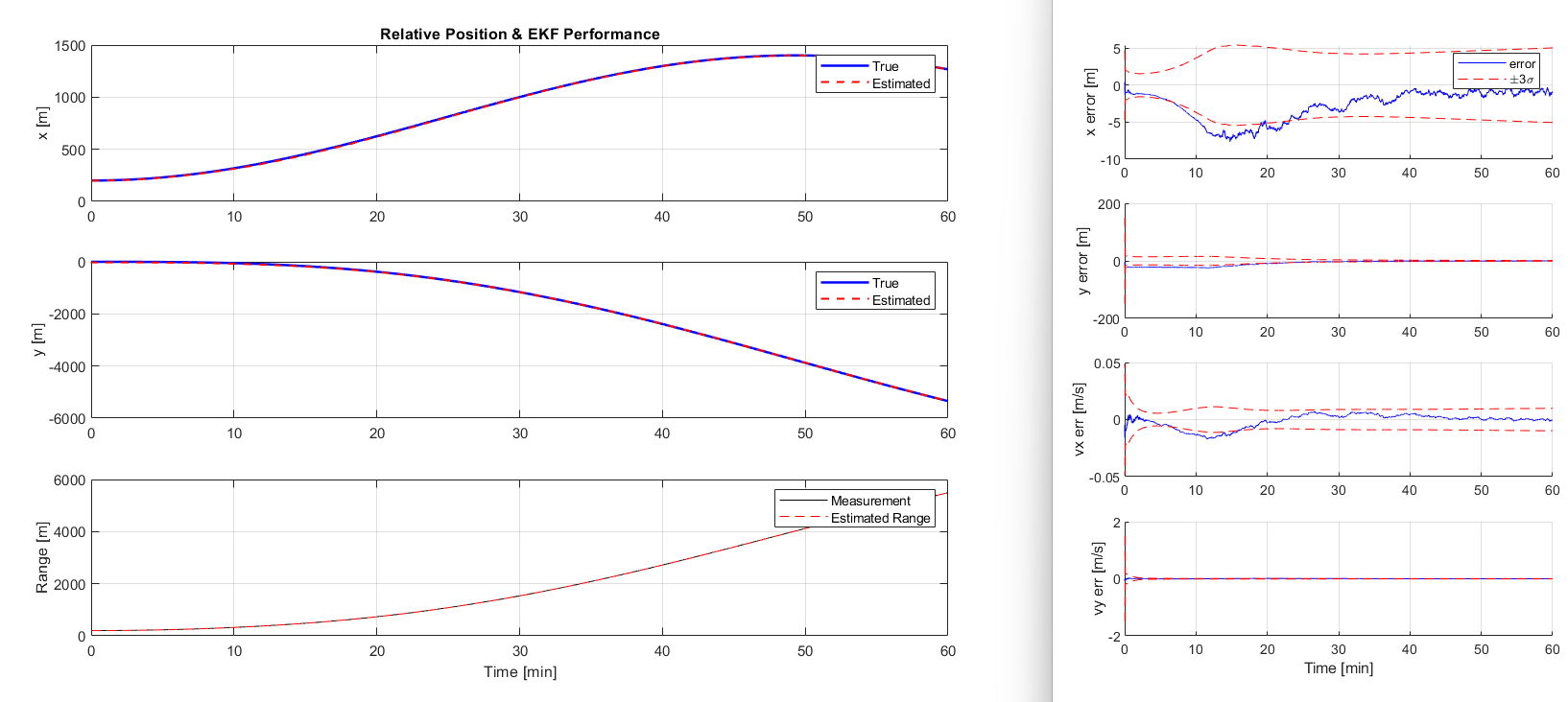}
        \caption{Instantaneous Throughput for Primary→Relay and Relay→Ground Links.}
        \label{nav-fig6}
\end{figure}

\begin{figure}
        \centering
        \includegraphics[width=0.5\linewidth]{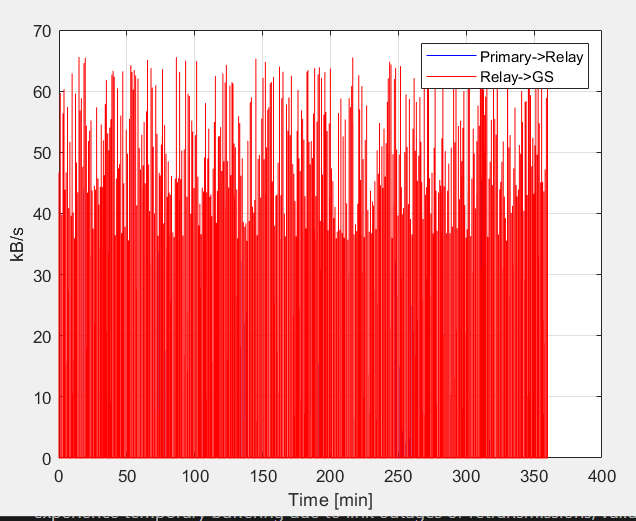}
        \caption{Throughput vs Time (Primary→Relay and Relay→GS).}
        \label{thrvstime-fig7}
\end{figure}

\begin{figure}
        \centering
        \includegraphics[width=0.5\linewidth]{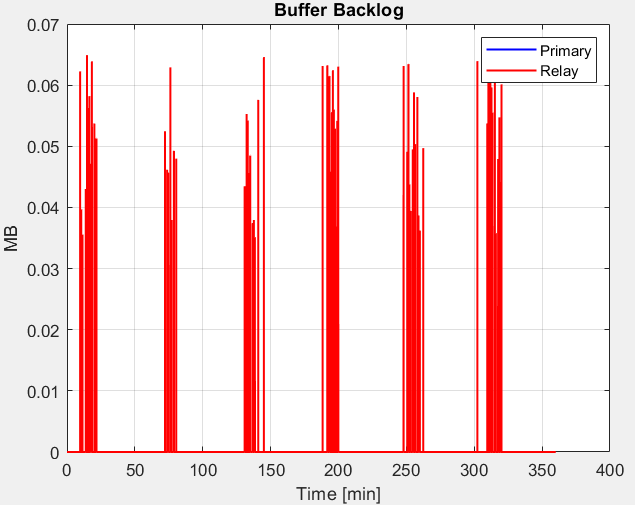}
        \caption{Buffer Backlog vs Time (Primary and Relay).}
        \label{buffer-fig9}
\end{figure}

The physical-layer performance of the communication architecture is highly dependent on orbital geometry and link availability. Fig.~\ref{linksnr-fig10} depicts the Signal-to-Noise Ratio (SNR) evolution, showing periodic oscillations for the Primary-to-Relay link (16--25~dB) and consistently stronger connectivity for the Relay-to-Ground link (30--38~dB). This SNR variation directly dictates the instantaneous throughput shown in Figs.~\ref{nav-fig6} and \ref{thrvstime-fig7}. The Relay-to-Ground link achieves peak rates between 35--65~kB/s, facilitating the rapid offloading of accumulated data, whereas the Primary-to-Relay link exhibits bursty, intermittent throughput. Fig.~\ref{buffer-fig9} confirms that the store-and-forward mechanism effectively manages these fluctuations; buffer backlogs at the relay show periodic spikes up to 0.065~MB during outages but drain rapidly upon contact, proving that the system capacity is sufficient to prevent long-term congestion or data loss.

\begin{figure}
    \centering
    \includegraphics[width=0.5\linewidth]{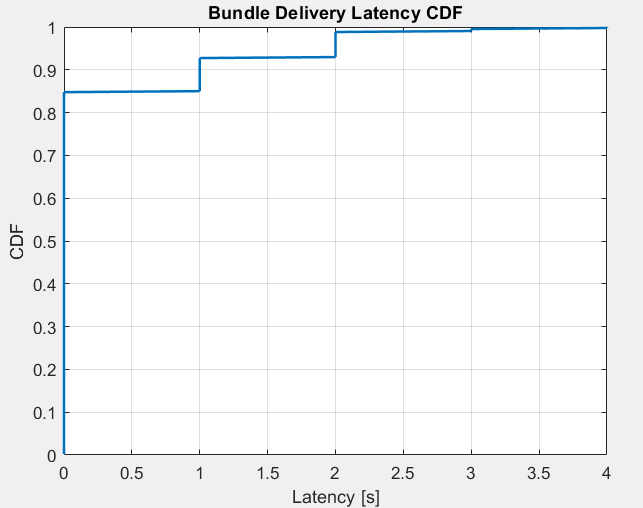}
    \caption{Bundle Delivery Latency CDF.}
    \label{bundledel-fig11}
\end{figure}

\begin{figure}
        \centering
        \includegraphics[width=0.5\linewidth]{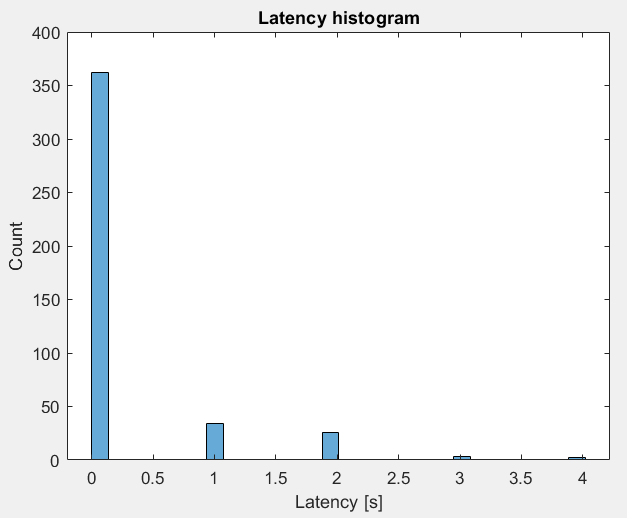}
        \caption{Latency Histogram (Bundle Delivery Delay).}
        \label{latency-fig8}
\end{figure}

\begin{figure}[!htbp]
    \centering
    \includegraphics[width=0.5\linewidth]{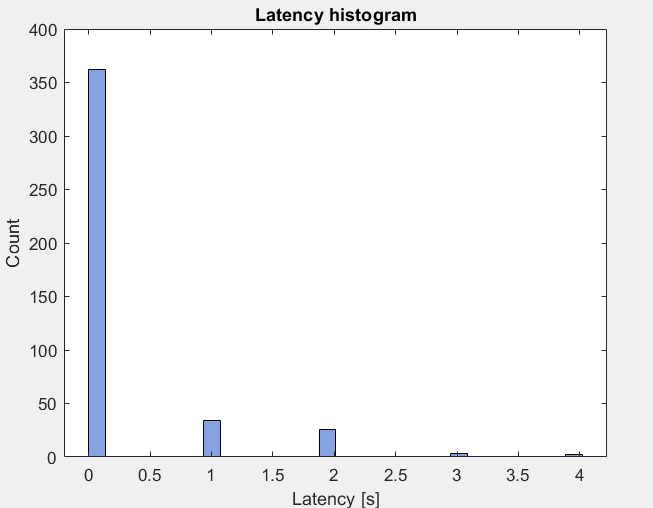}
    \caption{Latency Histogram.}
    \label{latencyhis-fig12}
\end{figure}

Finally, the efficiency of the Delay-Tolerant Networking (DTN) protocol was quantified through end-to-end latency analysis. The Cumulative Distribution Function (CDF) in Fig.~\ref{bundledel-fig11} reveals that approximately 90\% of bundles are delivered within 1 second, indicating highly efficient forwarding during simultaneous link availability. This is corroborated by the short-term latency histogram in Fig.~\ref{latency-fig8}, which shows a concentration of measurements below 0.2~s, with minor clustering around 1--2~s due to temporary buffering. 

However, the system's robustness under stress is highlighted in Fig.~\ref{latencyhis-fig12}, which captures the ``long tail'' of the distribution over a 6-hour simulation. Here, latencies range from 1,000~s to 4,000~s, corresponding to data generated during extended link outages. These results validate the architecture’s ability to handle both immediate telemetry and delay-tolerant bulk data without packet loss, confirming the operational viability of the proposed communication stack.

\section{Conclusion}\label{sec5}

% conclude results, briefly summarize, and future scope
% limitations of the study

This study validated a multi-debris-remediation architecture that integrates a NASA Evolutionary Xenon Thruster with mechanical clamping. GMAT and MATLAB simulations confirmed successful solar-electric deorbiting (800 km to 1000 km). The system achieved a position RMS error of $<10$ m using radar-based EKF navigation and a DTN delivery efficiency of 93\% within 1 s. However, gradual EKF divergence during continuous low-thrust phases, due to Clohessy-Wiltshire linearization errors, indicates a need for fidelity nonlinear modeling.

Future research aims to improve navigational robustness by replacing the EKF with an Unscented Kalman or particle filter to address nonlinear dynamics and state divergence. Validation will proceed to hardware-in-the-loop (HIL) testing of the clamping system and the edge-AI pipeline. Additionally, the framework will incorporate multi-agent reinforcement learning for trajectory planning and investigate hybrid propulsion for terminal-capture agility. Concurrently, we will extend reinforcement learning architectures to high-level mission planning to optimize autonomous task allocation and scheduling. Ultimately, these advancements aim to deploy autonomous fleets to mitigate Kessler Syndrome.

\section*{Code Availability}
The module implementation and experimentation code used in this study is available in the GitHub repository: \textit{\textbf{https://github.com/om520/Space-debri-removal-device}}

% \bibliographystyle{ieeetr}
% \bibliography{ref}

\end{document}